\documentclass[12pt,a4paper]{article}

\newcommand{\pa}{\partial}

\begin{document}

\begin{flushright}
hep-th/0110008
\end{flushright}
\vspace{1.8cm}

\begin{center}
 \textbf{\Large Nonstatic $AdS_2$ Branes and \\ 
the Isometry Group of $AdS_3$ Spacetime}
\end{center}
\vspace{1.6cm}
\begin{center}
 Shijong Ryang
\end{center}

\begin{center}
\textit{Department of Physics \\ Kyoto Prefectural University of Medicine
\\ Taishogun, Kyoto 603-8334 Japan}
\par
\texttt{ryang@koto.kpu-m.ac.jp}
\end{center}
\vspace{2.8cm}
\begin{abstract}
For the D-branes on the $SL(2,R)$ WZW model we present a particular
choice of outer automorphism for the gluing condition of currents
that leads to a special $AdS_2$ brane configuration. This 
configuration is shown to be a static solution in the cylindrical 
coordinates, and a nonstatic solution in the Poincar\'e coordinates
to the nonlinear equation of motion for the Dirac-Born-Infeld action
of a D-string. The generalization of it gives a family of nonstatic
$AdS_2$ brane solutions. They are deomonstrated to transform to each
other under the isometry group of $AdS_3$ spacetime.
\end{abstract}
\vspace{3cm}
\begin{flushleft}
September, 2001 
\end{flushleft}
\newpage
\section{Introduction}

Recently the D-branes on the group manifolds have attracted a great deal
of attention. The D-branes of the WZW models have been studied by using
the algebraic methods of the boundary conformal field theory or
the geometric methods where the geometries of the associated D-branes are
specified in terms of the gluing conditions for the currents \cite{KS,
KO,AS,SS,ARS,FFFS,SST,BDS,JP,AR,STA}. Owing to the conditions the 
endpoints of open strings are forced to stay on the (twined) conjugacy
classes of the group. For the compact $SU(2)$ group a D2-brane wrapped on
a two-sphere subspace of the three-sphere manifold has been shown to be 
stabilized against shrinking by the flux of worldvolume $U(1)$ gauge 
field \cite{BDS}. The stable spherical D2-brane configuration constructed
by the geometric and semiclassical anaysis to the Dirac-Born-Infeld 
(DBI) action can be identified with a Cardy state
which is a linear superposition of Ishibashi states in the boundary
conformal field theory. 

The D-branes on the noncompact group $SL(2,R)$ have been investigated 
by the geometric approach \cite{BP}. Gluing the currents by the inner
automorphisms leads to the regular conjugacy classes that are associated
with a two-dimensional hyperbolic plane ($H_2$), the light cone and a
de Sitter brane ($dS_2$) which describes a tachyonic unphysical 
D-string. On the other hand gluing the currents by an outer 
automorphism leads to the twined conjugacy class that is associated
with a two-dimensional anti de Sitter brane ($AdS_2$) which is a 
physical D-string stretched between two points on the boundary of
$AdS_3$. From the algebraic approach it has been indicated that the 
$dS_2$ brane corresponds to the continuous representation of $SL(2,R)$,
while the $H_2$ and $AdS_2$ branes correspond to the discrete 
representation \cite{GKS}. Further there have been various studies of the
classical motions and the spectrum of open strings attached to the
$AdS_2$ branes \cite{PR,LOPT,LL} and the constructions of the Cardy 
states based on the discrete \cite{HS} and continuous representations
\cite{RR}.

In Ref. \cite{BP} the static $AdS_2$ brane configuration corresponding to
a particular outer automorphism was constructed in the $SL(2,R)$ WZW
model. In order to try to generalize this special $AdS_2$ brane on 
the $SL(2,R)$ manifold we will propose the other type of 
outer automorphism and show how the static or nonstatic D-brane 
configuration is described by the twined conjugacy class associated
with this type of automorphism and satisfies the equation of motion for
the DBI action of the D-string accompanied with the $U(1)$ gauge field.
The generalization of these automorphisms is performed and the 
general time-dependent solutions are constructed. These stationary
configurations are shown to be related to each other by the isometry
group of the $AdS_3$ spacetime.

\section{General $AdS_2$ branes}

We consider the D-branes in the $AdS_3$ spacetime with the radius $L$ that
is isomorphic to the group manifold of $SL(2,R)$ and is represented by the
hyperboloid
\begin{equation}
 - (X^0)^2 - (X^3)^2 + (X^1)^2 + (X^2)^2 = - L^2
\label{hyp}\end{equation}
embedded in $\mathrm{R}^{2,2}$. A general group element 
can be parametrized as
\begin{eqnarray}
 g = \frac{1}{L} \left( \begin{array}{cc} X^0 + X^1 & X^2 + X^3 \\
 X^2 - X^3 & X^0 -X^1 \end{array}  \right),
\end{eqnarray}
whose determinant is equal to one through  (\ref{hyp}). The metric on
$AdS_3$, $ds^2 = -(dX^0)^2 - (dX^3)^2 + (dX^1)^2 + (dX^2)^2$ is 
expressed in the cylindrical coordinates $(\tau, \rho, \phi)$
\begin{equation}
 X^0 + iX^3 = L \cosh \rho e^{i\tau}, \;  X^1 + iX^2 = L \sinh 
 \rho e^{i\phi}
\end{equation}
as
\begin{equation}
ds^2 = L^2(- \cosh^2\rho d\tau^2 + d\rho^2 + \sinh^2\rho d\phi^2 )
\end{equation}
and the Neveu-Schwarz antisymmetric tensor is given by
\begin{equation}
H = dB = L^2\sinh(2\rho)d\rho\wedge d\phi\wedge d\tau.
\label{nsh}\end{equation}
The radial coordinate $\rho$, the angular coordinate $\phi$ and the 
global time coordinate $\tau$ range over $0\le \rho < \infty, 
0\le \phi < 2\pi$ and $-\infty < \tau < \infty$, where the boundary
of $AdS_3$ is at $\rho \rightarrow \infty$. The radius of $AdS_3$ is
specified by the level of the $SL(2,R)$ current algebra as
$L^2 \sim |k|\alpha'$ in the semiclassical limit.  Alternatively 
there are the Poincar\'e coordinates $(t, x, u)$ defined by
\begin{equation}
X^0 + X^1 = Lu,\; X^2 \pm X^3 = Luw^{\pm}, \; X^0 - X^1 
= L \left( \frac{1}{u} + uw^+ w^- \right)
\end{equation}
with $w^{\pm} \equiv x \pm t$. In these coordinates that range over the
entire  $\mathrm{R}^3$, the metric is expressed as
\begin{equation}
ds^2 = L^2 \left( \frac{du^2}{u^2} + u^2 dw^+ dw^- \right),
\label{mep}\end{equation}
where the boundary of $AdS_3$ is specified by $|u| \rightarrow \infty$.
From the gluing condition $J = R\bar{J}$ identifying the left and right
moving $SL(2,R)$ currents of the WZW model modulo the Lie algebra 
automorphism $R$, the worldvolumes of D-branes containing a fixed 
element $g$ of the group are represented by the ( twined ) conjugacy 
classes 
\begin{equation}
\mathcal{W}^{\omega}_g = \{ \omega(h)gh, {}^{\forall}h \in SL(2,R) \},
\end{equation}
where $\omega$ is a group automorphism induced near the identity from
$R$ \cite{AS,FFFS,SST}. For the inner automorphism of $SL(2,R)$, 
$\omega(h) = g^{-1}_0 h g_0$ with $g_0 \in SL(2,R)$ the worldvolumes 
of D-branes are characterized by the regular conjugacy classes and
expressed as $H_2$, the light cone and $dS_2$. When $\omega$ is a
nontrivial outer automorphism
\begin{equation}
\omega(h) = \omega^{-1}_0 h \omega_0, \; \mathrm{with} \;
\omega_0 = \left( \begin{array}{cc} 0 & 1 \\ 1 & 0 \end{array} \right),
\label{omf}\end{equation}
where $\omega_0$ has minus one determinant so as not to be an element 
of $SL(2,R)$, the D-brane worldvolume 
lies on the twined conjugacy class
\begin{equation}
 \mathrm{tr}(\omega_0 g) = \frac{2X^2}{L} = 2C
\label{tcs}\end{equation}
for some constant $C$ and describes the $AdS_2$ geometry, $(X^0)^2 + 
(X^3)^2 - (X^1)^2 = L^2( 1 + C^2 )$ which is embedded in $AdS_3$.
In Ref. \cite{BP} from the DBI action of a D-string the static $AdS_2$
brane solution was derived by analysing the continuity equation of 
the energy-momentum tensor or minimizing the energy of the static 
state in the Poincar\'e coordinates.

Instead of the particular outer automorphism (\ref{omf}) we first 
consider another choice 
\begin{equation}
\omega_0 = \left( \begin{array}{cc} 1 & 0 \\ 0 & -1 \end{array}
\right),
\label{omt}\end{equation}
whose determinant is minus one and manipulate directly the
equation of motion for the DBI action. In this choice the outer 
automorphism is the operation that changes the sign of $X^2$ and 
$X^3$, while leaving $X^0$ and $X^1$ unchanged. The twined conjugacy
class given by
\begin{equation}
 \mathrm{tr}(\omega_0g) = \frac{2X^1}{L} = 2C
\label{tcf}\end{equation}
with some constant $C$ yields the $AdS_2$ geometry expressed  
$(X^0)^2 + (X^3)^2 - (X^2)^2 = L^2( 1 + C^2 )$. The DBI action 
of a D-string with tension $T_D$ is given by
\begin{equation}
I = \int d^2\sigma \mathcal{L} = -T_D \int d^2\sigma \sqrt{-\det
(\hat{g} + \hat{B} + 2\pi \alpha'F)},
\end{equation}
where $\hat{g}$ and $\hat{B}$ are the pullbacks of the WZW backgrounds
to the D-string worldvolume and $F$ is the worldvolume $U(1)$ electric
field. In the cylindrical coordinates from (\ref{tcf}) the $AdS_2$ brane
is described by 
\begin{equation}
\sinh \rho \cos \phi= C.
\label{coc}\end{equation}
This $AdS_2$ brane is static as embedded by $\rho = \sinh^{-1}
(C/\cos\phi)$. Therefore the static gauge $\sigma^0 = \tau, \sigma^1 =
\phi$ forces the DBI action to be 
\begin{equation}
I = -T_D \int d\tau d\phi\sqrt{L^4\cosh^2\rho(\sinh^2\rho +
(\pa_{\phi}\rho)^2) - \mathcal{F}^2_{\phi\tau}}.
\end{equation}
From (\ref{nsh}) we take a convenient gauge for the Neveu-Schwarz field
to have $B = L^2\sinh^2\rho d\phi\wedge d\tau$ so that the invariant
combination  $\mathcal{F}_{\phi\tau}$ is given by 
$\mathcal{F}_{\phi\tau} = L^2\sinh^2\rho - 2\pi\alpha' \dot{A}_{\phi}$.
Since the Wilson line $A_{\phi}$ is a cyclic variable, its conjugate 
momentum is a quantized constant of motion
\begin{equation}
\frac{1}{2\pi}\Pi_{\phi} \equiv \frac{\pa \mathcal{L}}{\pa 
\dot{A}_{\phi}} = - \frac{2\pi\alpha' T_D \mathcal{F}_{\phi\tau}}
{L^2 \sqrt{D}} = -q \in \mathbf{Z},
\label{pi}\end{equation}
where $D = \cosh^2\rho( \sinh^2\rho + (\pa_{\phi}\rho)^2) - 
\mathcal{F}^2_{\phi\tau}/L^4$. The integer $q$ expresses the number of
fundamental strings bound to the D-string, since an electric field on
the D-string is equivalent to a fundamental string.  The equation of 
motion for $\rho$ is given by
\begin{eqnarray}
\frac{\sinh^2\rho + \cosh^2\rho + (\pa_{\phi}\rho)^2}{\sqrt{D}} &-& 
\frac{1}{\sinh\rho\cosh\rho} \pa_{\phi}\left( \frac{\cosh^2\rho\pa_{\phi}
\rho}{\sqrt{D}} \right) \nonumber \\
 &=& \frac{2(\sinh^2\rho + f)}{\sqrt{D}},
\label{eqr}\end{eqnarray}
where $f \equiv  2\pi\alpha' F_{\phi\tau}/L^2$. Although we
must solve simultaneously the nonlinear equations (\ref{pi}) and 
(\ref{eqr}), here we will seek a condition for the configuration 
(\ref{coc}) to satisfy them. Through (\ref{pi}) the right hand side of 
the formidable-looking equation (\ref{eqr}) is written in terms 
of the tension $T_F$ of the fundamental string as $2qT_F/T_D$. Moreover,
from (\ref{pi}) $D$ is tautologically expressed as 
\begin{equation}
D = \frac{1}{1 + (\frac{qT_F}{T_D})^2} D_0 
\end{equation}
with $D_0 = \cosh^2\rho \sinh^2\rho + \cosh^2\rho(\pa_{\phi}\rho)^2$.
The substitution of (\ref{coc}) into $D_0$ yields $D_0 =
(C^4 + C^2)/\cos^4\phi$. This simplified expression can then be used
to derive
\begin{equation}
\frac{C^2\sqrt{1 + (\frac{qT_F}{T_D})^2}}{\cos^2\phi \sqrt{D_0}} =
\frac{qT_F}{T_D},
\end{equation}
when (\ref{coc}) is substituted into the nonlinear equation (\ref{eqr}).
As long as $q \ge 0$, this equation can produce a solution 
$C = \pm qT_F/T_D$. We have demonstrated
that the $AdS_2$ static configuration (\ref{coc}) indeed satisfies the
stationary equation for the DBI action of a D-string only when $C$
takes the special value in proportion to $q$. 

For the twined conjugacy class (\ref{tcs}) the $AdS_2$ static 
configuration given by 
\begin{equation}
\sinh\rho \sin\phi = C 
\label{sic}\end{equation}
satisfies the stationary equation only when $C = \pm qT_F/T_D$ that is
obtained from 
\begin{equation}
\frac{C^2\sqrt{1 + (\frac{qT_F}{T_D})^2}}{\sin^2\phi \sqrt{D_0}} =
\frac{qT_F}{T_D}
\end{equation}
with $D_0 = (C^4 + C^2)/\sin^4\phi$. The obtained $C$ in the cylindrical
coordinates agrees with $C$ for the static $AdS_2$ brane solution
$u = C/x$ in the Poincar\'e coordinates \cite{BP}.

Here we turn to the Poincar\'e coordinates. For the twined conjugacy 
class (\ref{tcf}) the $AdS_2$ brane configuration is specified by
\begin{equation}
\left( \frac{1}{u} + C \right)^2 + x^2 - t^2 = 1 + C^2. 
\end{equation}
From it $u$ is so expressed in terms of $t$ and $x$ as
\begin{equation}
u = \frac{1}{\pm \sqrt{A} - C}, \; \mathrm{with}\; A \equiv t^2 - x^2
+ C^2 + 1
\label{sou}\end{equation}
that the $AdS_2$ brane is nonstatic. We choose a gauge for the 
Neveu-Schwarz potential to be $B = L^2u^2dx\wedge dt$. Hence the gauge
invariant two-form is given by $\mathcal{F} = L^2(u^2 + f)dx\wedge dt$
with $f = 2\pi\alpha' F_{xt}/L^2$. A static gauge $\sigma^0 = t,
\sigma^1 = x$ is taken so that the DBI action for a D-string is
expressed as
\begin{equation}
I = - T_D L^2\int dtdx \sqrt{D},
\end{equation}
where $D = u^4 + (\pa_xu)^2 - (\pa_tu)^2 - (u^2 + f)^2$. The equation of
motion for the $U(1)$ gauge field gives the Gauss constraint
\begin{equation}
\frac{2\pi\alpha' T_D \mathcal{F}_{xt}}{L^2 \sqrt{D}} = q \in \mathbf{Z},
\label{gau}\end{equation}
where $q$ is the number of electric flux quanta turned on along the
$AdS_2$. On the other hand the equation of motion for $u$ is provided by
\begin{equation}
\frac{2u^3}{\sqrt{D}} + \pa_t\left(\frac{\pa_tu}{\sqrt{D}}\right) -
\pa_x\left(\frac{\pa_xu}{\sqrt{D}}\right) = \frac{2u(u^2 + f)}{\sqrt{D}},
\label{equ}\end{equation}
whose right hand side simplifies to $2uqT_F/T_D$ due to the Gauss
constraint (\ref{gau}). From (\ref{gau}) $D$ is also equivalently 
rewritten by
\begin{equation}
D = \frac{1}{1 + (\frac{qT_F}{T_D})^2} D_0,\; \mathrm{with}\;
D_0 =  u^4 + (\pa_xu)^2 - (\pa_tu)^2. 
\label{do}\end{equation}
Plugging the nonstatic configuration (\ref{sou}) into $D_0$ we have also
a simplified expression $D_0 = (1 + C^2)/A(\pm \sqrt{A}- C)^4$.
Owing to this expression the nonlinear equation (\ref{equ}) into which
(\ref{sou}) is substituted, becomes to take the compact form
\begin{equation}
\pm \frac{2Cu\sqrt{1 + (\frac{qT_F}{T_D})^2}}{\sqrt{1 + C^2}}
= \frac{2uqT_F}{T_D}.
\label{pme}\end{equation}
In view of this expression we extract $C = \pm qT_F /T_D$ whose sign
corresponds to the the sign of (\ref{pme}). Thus we have an $AdS_2$
brane classical solution in the Poincar\'e coordinates
\begin{equation}
u = \pm \frac{1}{\sqrt{t^2 - x^2 + (\frac{qT_F}{T_D})^2 + 1} - 
\frac{qT_F}{T_D}},
\label{upm}\end{equation}
which exhibits the motion of one D-string bound to $q$ fundamental 
strings.

Now we consider the other choice for the outer automorphism
\begin{equation}
w_0 = \left( \begin{array}{cc} \sin\theta & \cos\theta \\
\cos\theta & -\sin\theta \end{array} \right),
\label{wsi}\end{equation}
which interpolates between $w_0 = \sigma_1$ in (\ref{omf}) and
$w_0 = \sigma_3$ in (\ref{omt}) where $\sigma_1$ and $\sigma_3$ are
the Pauli matrices. The twined conjugacy class characterized by
\begin{equation}
\mathrm{tr}(w_0g) = \frac{2(\sin \theta X^1 + \cos \theta X^2)}{L}
= 2C
\label{trw}\end{equation}
specifies the D-brane worldvolume which is expressed as the hyper surface
\begin{equation}
(X^0)^2 + (X^3)^2 - (\sec\theta X^1 - LC\tan\theta )^2 = L^2( 1 + C^2 )
\end{equation}
for $\cos\theta \neq 0$ or 
\begin{equation}
(X^0)^2 + (X^3)^2 - (\mathrm{cosec}\theta X^2 
- LC\cot\theta )^2 = L^2( 1 + C^2 )
\end{equation}
for $\sin\theta \neq 0$. In the cylindrical coordinates the 
Eq. (\ref{trw}) reads
\begin{equation}
\sinh\rho \sin(\phi + \theta) = C,
\end{equation}
which is generated by making a rotation of the $\phi$-direction by
$\theta$ from the $AdS_2$ solution (\ref{sic}) associated with $\omega_0
= \sigma_1$. If we make the following choice for $\omega_0$
\begin{equation}
\omega_0 = \left( \begin{array}{cc} \sinh\varphi & \cosh\varphi \\
\cosh\varphi & \sinh\varphi \end{array} \right),
\label{wsh}\end{equation}
which is not connected to (\ref{omt}) but reduced to (\ref{omf})
at $\varphi = 0$, the worldvolume geometry of D-brane is described by
\begin{equation}
\sinh\varphi\cosh\rho\cos\tau + \cosh\varphi\sinh\rho\sin\phi = C.
\end{equation}
Since in the cylindrical coordinates $\rho$ is the involved function of
$\tau$ and $\phi$, we will analyse this type of configuration in the 
Poincar\'e coordinates by making the more general parametrization.

Let us consider the following general outer automorphism expressed 
in terms of four real parameters as 
\begin{equation}
\omega_0 = \left( \begin{array}{cc} \alpha & \beta \\ \gamma & \delta
\end{array} \right), \; \mathrm{with} \;\alpha\delta - \beta\gamma = -1,
\label{gwo}\end{equation}
which includes (\ref{wsi}) and (\ref{wsh}) as special cases. 
The worldvolume of D-brane lying on the twined conjugacy class 
$\mathrm{tr}(w_0g) = 2C$ in the Poincar\'e coordinates is given by
\begin{equation}
u = \frac{-\delta}{\pm\sqrt{A} - C}
\label{gu}\end{equation}
with $A \equiv C^2 - \delta(\alpha + \beta w^- +\gamma w^+ + 
\delta w^+w^- )$. In order to see that this general trajectory is indeed
the solution of the stationary nonlinear equation for the DBI action,
we calculate $D_0$ in (\ref{do}) similarly as
\begin{equation}
D_0 = \frac{\delta^4( C^2 + \beta\gamma - \alpha\delta )}
{A(\pm\sqrt{A} - C)^4},
\end{equation}
where remarkable cancellations happen without using $\alpha\delta - 
\beta\gamma = - 1$ and obtain a simple equation for $C$ from (\ref{equ})
\begin{equation}
\pm \frac{C\sqrt{1 + (\frac{qT_F}{T_D})^2}}{\sqrt{\beta\gamma - 
\alpha\delta + C^2}} = \frac{qT_F}{T_D}.
\label{cqt}\end{equation}
For $\alpha\delta - \beta\gamma = -1$, the constant $C$ is so evaluated
as $\pm qT_F/T_D$ again that we have a general class of 
nonstatic solutions
\begin{equation}
u = \pm \frac{-\delta}{\sqrt{(\frac{qT_F}{T_D})^2 - \delta(\alpha + 
\beta w^- +\gamma w^+ + \delta w^+w^- )} - \frac{qT_F}{T_D} }.
\label{gso}\end{equation}
If we make a choice  $\alpha\delta - \beta\gamma = 1$ instead, the D-brane
worldvolume belongs to the regular conjugacy class. From (\ref{cqt}) 
we derive $C^2 = - (qT_F/T_D)^2$ which implies that $q$ becomes 
imaginary and this solution represents an unphysical brane with the 
supercritical electric field in the same way as the $dS_2$ brane.
From the general expression (\ref{gso}) the stationary solutions for
the choices (\ref{wsi}) and (\ref{wsh}) are respectively given by
\begin{eqnarray}
u &=& \pm \frac{1}{\sqrt{t^2 - (x -\cot\theta)^2 + (1+C_0^2)
\mathrm{cosec}^2\theta} -C_0\mathrm{cosec}\theta}, \;
\mathrm{for} \; \sin \theta > 0, \nonumber \\
u &=& \pm \frac{-1}{\sqrt{t^2 - (x +\coth\varphi)^2 + (1+C_0^2)
\mathrm{cosech}^2\varphi} -C_0\mathrm{cosech}\varphi},\;
\mathrm{for} \; \varphi > 0
\end{eqnarray}
with $C_0 \equiv qT_F/T_D$. In the limits $\theta \rightarrow 0$ and
$\varphi \rightarrow 0$ these time-dependent solutions 
are confirmed to reduce to the same static solution 
$u = \pm C_0/x$ that is associated with 
$\omega_0 = \sigma_1$. If we consider the more general case that 
$\omega_0$ is parametrized as (\ref{gwo}) but with 
$\alpha\delta - \beta\gamma = -z < 0$ for $z \neq 1$, $C$ is obtained
by $C = \pm C_0\sqrt{z}$ from (\ref{cqt}). Hence we have the general
solutions also expressed by (\ref{gso}) where $qT_F/T_D$ is replaced by
$\sqrt{z}qT_F/T_D$. They agree with the solutions obtained by starting
with the parametrization 
\begin{equation}
\omega_0 = \frac{1}{\sqrt{z}} \left(\begin{array}{cc} \alpha & \beta 
\\ \gamma & \delta \end{array} \right),
\end{equation} 
whose determinant is minus one. Therefore it is sufficient to restrict
ourselves to the $\det \omega_0 = -1$ sector. When we combine
a special outer automorphism
$\omega_0 = \sigma_1$ in (\ref{omf}) with the general inner automorphism
specified by $h$ we have $\mathrm{tr}(\omega_0hgh^{-1}) = 2C$ that 
describes the D-brane worldvolume chracterized by a combined automorphism
$\omega'_0 = h^{-1}\sigma_1h$. The general parametrization of $h$
\begin{equation}
h = \left( \begin{array}{cc} p & q \\ r & s \end{array} \right), \;
\mathrm{with} \; ps - qr = 1
\end {equation}
leads to
\begin{equation}
\omega'_0 = \left( \begin{array}{cc} rs - pq & s^2 - q^2 \\
p^2 - r^2 & -( rs - pq ) \end{array} \right),
\label{woi}\end{equation}
which povides some class of outer automorphism since its determinant
is minus one and its trace is zero. The outer automorphisms such as
(\ref{omt}) and (\ref{wsi}) belong to the class of (\ref{woi}) but
the other outer automorphism (\ref{wsh}) does not belong to it.

\section{$SL(2,R)_L \times SL(2,R)_R$ transformations}

We elucidate the relations among the above solutions. The metric 
(\ref{mep}) in the Poincar\'e coordinates
 is rewritten in terms of $y = 1/u$ as
\begin{equation}
ds^2 = \frac{L^2}{y^2}( dy^2 + dw^+dw^- ),
\label{mey}\end{equation}
where the $AdS_3$ boundary is at $y = 0$. In Ref. \cite{MS} it was
presented that the $AdS_3$ metric (\ref{mey}) has the following 
$SL(2,R)_L \times SL(2,R)_R$ isometry group. The $SL(2,R)_L$ 
transformation is given by
\begin{eqnarray}
w^+ & \rightarrow & {w^+}' = \frac{aw^+ + b}{cw^+ + d}, \; w^- \rightarrow
{w^-}' = w^- + \frac{cy^2}{cw^+ + d}, \nonumber \\
y & \rightarrow & y' = \frac{y}{cw^+ + d}
\end{eqnarray}
with real $a, b, c, d$ obeying $ad - bc = 1$, while the $SL(2,R)_R$ 
transformation is
\begin{eqnarray}
w^+ & \rightarrow & {w^+}' = w^+ + \frac{cy^2}{cw^- + d}, \;
w^-  \rightarrow  {w^-}' = \frac{aw^- + b}{cw^- + d}, \nonumber \\
y & \rightarrow & y' = \frac{y}{cw^- + d}.
\end{eqnarray}
Both the transformations map the $AdS_3$ boundary at $y = 0$ to itself
and act on the boundary as the usual conformal transformations of 
$1+1$ dimensional Minkowski spacetime. In the $SL(2,R)_L$ transformation
the actions on $y$ and $w^-$ are translated into 
$u \rightarrow u' = u(cw^+ + d)$ and $w^- \rightarrow {w^-}' = 
w^- + c/u^2(cw^+ + d)$ for the metric (\ref{mep}), 
while the action on $w^+$ is not altered. 
Under the $SL(2,R)_L$ transformation the static solution of the form
$u = C/x = 2C/(w^+ + w^-)$ for (\ref{tcs}) corresponding to the choice
$\omega_0 = \sigma_1$ in (\ref{omf}) is mapped to
\begin{equation}
u' = \frac{2C}{(a - c{w^+}'){w^-}' - ( b - d{w^+}') - \frac{c}{u'^2}},
\end{equation}
which reads
\begin{equation}
u' = \frac{c}{\pm \sqrt{C^2 + c(-b + a{w^-}' + d{w^+}' - c{w^+}'{w^-}')}
 - C}.
\label{tul}\end{equation}
Under the $SL(2,R)_R$ transformation with $u \rightarrow u' = 
u(cw^- + d)$ the static solution is similarly mapped to
\begin{equation}
u' = \frac{c}{\pm \sqrt{C^2 + c(-b + d{w^-}' + a{w^+}' - c{w^+}'{w^-}')}
 - C}.
\label{tur}\end{equation}
If we choose 
\begin{equation}
\left( \begin{array}{cc} a & b \\ c & d \end{array} \right) =
\left( \begin{array}{cc} 0 & -1 \\ 1 & 0 \end{array} \right),
\end{equation}
whose determinant is equal to one, both the transformed solutions 
(\ref{tul}), (\ref{tur}) are identical to the nonstatic solution of the 
form (\ref{sou}) corresponding to the choice $\omega_0 = \sigma_3$ 
in (\ref{omt}). It is interesting to note that the $SL(2,R)_L$ 
transformed solution (\ref{tul}) is just identical to the general form of
solution (\ref{gu}), if we choose
\begin{equation}
\left( \begin{array}{cc} a & b \\ c & d \end{array} \right) =
\left( \begin{array}{cc} \beta & -\alpha \\ -\delta & \gamma \end{array}
\right),
\end{equation}
whose determinant is equal to one through the relation $\alpha\delta - 
\beta\gamma = -1$ in (\ref{gwo}). In the same way the $SL(2,R)_R$ 
transformed solution (\ref{tur}) also agrees with the general form of 
solution (\ref{gu}), when the transformation parameters are chosen as 
\begin{equation}
\left( \begin{array}{cc} a & b \\ c & d \end{array} \right) =
\left( \begin{array}{cc} \gamma & -\alpha \\ -\delta & \beta \end{array}
\right),
\end{equation}
which has unit determinant. Thus there are nontrivial $SL(2,R)_L \times
SL(2,R)_R$ actions on the $\omega_0 = \sigma_1$ solution that give not
only the $\omega_0 = \sigma_3$ solution but also the general class of
solutions. 

\section{Conclusions}

We have observed that in the $SL(2,R)$ WZW model 
 the $AdS_2$ brane configuration specified by an
outer automorphism $\omega_0=\sigma_3$ is static in the cylindrical
coordinates and nonstatic in the Poincar\'e coordinates, which is 
compared to the $AdS_2$ brane configuration
specified by  $\omega_0=\sigma_1$ that is static
in both the coordinates. By manipulating the DBI action of a D-string
carrying the $U(1)$ electric field in the $AdS_3$ spacetime we have
demonstrated that these static and nonstatic configurations satisfy
amazingly the involved nonlinear equation of motion, only when the
parameter $C$ chracterizing the shape of each trajectory takes the
same quantized value. These classical solutions are labelled by the
integer $q$, the number of fundamental strings which provides
the quantized $C$. From the general outer
automorphism we have found a family of 
time-dependent analytic solutions to the formidable-looking DBI
equation. As far as the determinant of $\omega_0$ that specifies the
general outer automorphism is minus one, the parameter $C$ associated
with the general class of nonstatic solutions is independent of the 
parametrization of $\omega_0$. This general outer automorphism is
more wide class than the automorphism generated by combining a special
outer automorphism specified by $\omega_0 = \sigma_1$ and the general
inner automorphism. We have shown that the nontrivial static and
nonstatic configurations are transformed to each other by the isometry
group $SL(2,R)_L \times SL(2,R)_R$ of $AdS_3$ spacetime in the 
Poincar\'e coordinates.

Our demonstration of the nontrivial solutions is expected to give a
clue to the construction of the similar time-dependent classical
analytic solutions for the DBI action of the low-dimensional
D-branes such as D-string and D2-brane 
with the $U(1)$ gauge field in the noncompact part $AdS_{p+2}$
of the near horizon geometry created by the
D$p$-branes in the string theory. The $U(1)$ electric field may
play an important role for the existence of such classical solutions
and simplification of the nonlinear DBI equation of motion.

\end{document}